\documentstyle[manuscript,prd,aps]{revtex}
\begin{document}
\tighten
\draft
\title{SURFACE CURRENT-CARRYING DOMAIN WALLS}
\author{Patrick PETER}
\address{D\'epartement d'Astrophysique Relativiste et de Cosmologie,\\
Observatoire de Paris-Meudon, UPR 176, CNRS, 92195 Meudon (France),\\
Department of Applied Mathematics and Theoretical Physics,\\
University of Cambridge, Silver Street, Cambridge CB3~9EW (England),\\
{\rm and}\\ Isaac Newton Institute for Mathematical Sciences,\\
20 Clarkson Road, Cambridge CB3 0EH (England).}
\date{March 1995}
\preprint{\vbox{ \hbox{DAMTP-R95/6} \hbox{hep-ph/9503408}}}
\maketitle
\begin{abstract}
Domain walls, arising from the spontaneous breaking of a discrete
symmetry, can be coupled to charge carriers. In much the same way as
the Witten model for superconducting cosmic string, an investigation
is made here in the case of $U(1)\times Z_2 \to U(1)$, where a bosonic
charge carrier is directly coupled to the wall-forming Higgs field.
All internal quantities, such as the energy per unit surface and the
surface current, are calculated numerically to provide the first
complete analysis of the internal structure of a surface
current-carrying domain wall.
\end{abstract}

\section{Introduction}

Domain walls~\cite{kibble,book} can arise in various grand unified
theories (GUT) whenever a discrete symmetry is broken by means of a
Higgs field. Because they have immediately been shown to induce a
cosmological catastrophe~\cite{kibble} even if they appear in a very
late phase transition, their internal structure has not been studied
in much details yet, since it was widely believed that they could not
survive until now. Indeed, with an energy per unit surface of the
order of the cube of the symmetry breaking energy scale $\eta$ say, a
single wall crossing the universe, even with $\eta$ as low as
$\eta\sim 100$~GeV, would produce an enormous overdensity $\Omega
_{\rm Wall} \sim 10^8$, or, in the case where only small balls were
to survive, very large anisotropies in the cosmic microwave background
radiation would be induced which are not observed~\cite{book}. Hence,
if stable walls are to exist in a theory, one must have an inflationary
period between the time they were formed and now.

The general belief nowadays concerning domain walls, assuming they
were ever produced at all, is examplified by the Peccei-Quinn phase
transition~\cite{PQ}, whose cosmological relevance notably for the
axion problem is still the subject of open
discussion~\cite{discuss}. The idea is that even though walls
could have been formed, the corresponding phase transition would have
been preceded by a string forming transition in such a way that domain
walls could only be bounded by strings. In such a framework, all walls
would have had a finite size, huge surface tension, and would have
evaporated in less than a Hubble time, thereby effectively solving the
problem. It could therefore appear that studying their internal
structure is indeed pointless.

However, just like in the case of cosmic strings, the situation could
be rather different if domain walls were to have the ability to carry
some sort of charge. In the case of strings, a current has the
immediate effect of breaking the Lorentz symmetry along the
worldsheet, so that one can consider rotating loops (called
vortons~\cite{vorton} because of their particle-like properties, or
rings~\cite{ring}). The point is that cosmic strings are believed to
scale (see Ref.~\cite{book} for a recent review) because
the network of string is dominated by the loops, who eventually
gravitationally radiate all their energy away. When a current is
present, these loops might reach equilibrium
configurations~\cite{vorton,ring} whose classical stability was
recently discussed~\cite{stabgen,stabwit} with the result that if no
quantum instability exists, the scaling is spoiled and they could
easily overfill the universe unless they were produced at a very low
energy scale (estimated at $\sim 10$~TeV).

Now if the strings bounding the walls were superconducting, the
problem could in fact be rather similar, the presence of a domain wall
modifying the equilibrium configuration in an unknown way, while
presumably modifying the constraint. This issue,
which can, and should be analysed in the framework of Carter's
formalism~\cite{formal} for describing $p$-branes, is still a
completely open subject. Another difficulty can arise
in the case where strings are not current-carrying, but if
the wall itself is. Indeed, the point is, as before for the
cosmic string scenario, that the breaking of the Lorentz symmetry
along the worldsheet, whatever its intrinsic dimension, allows a
definition of rotation, and eventually the recognition of the
existence of centrifugally supported states. Of course, it is not clear
yet whether these objects could be formed and reach stable states
at all, and therefore their cosmological relevance has not
been established. However, in the purpose of studying these
frisbee-like configurations, it is necessary that one knows
the relevant internal quantities such as the energy per unit
area and the surface tensions: they are explicitely calculated
in the present article.

It may seem that coupling charged (or hypercharged) particles to a
domain wall forming Higgs fields is a bit arbitrary, but in view of
the fact that most topological defects are predicted to form in
various GUT models where the number of degrees of freedom, including
scalar, vector and fermion fields is huge, and where the couplings are
almost unrestricted, it seems fairly plausible. The purpose of this
article is thus to present a toy model, similar to the Witten bosonic
model~\cite{witten} for superconducting cosmic strings, where the
symmetry breaking scheme is simply $U(1)\times Z_2 \to U(1)$. This
model, much like the Witten model, is expected to yield qualitatively
relevant results. The work is arranged as follows: after presenting the
actual model in a first section, we investigate the microscopic
structure of such a wall and end up by dealing with the abovementionned
integrated internal quantities, namely the energy per unit area,
the surface tensions as well as the surface current. The equation of
state, relating these quantities, is then computed numerically
from the solution of the field equations and is shown to share
most of the superconducting cosmic string equation of
state properties~\cite{neutral}, and in particular the existence of
a phase frequency threshold, which is discussed in some lenght at
the end of the paper. This study of a domain wall model thus shed a new
light on the general knowledge on current-carrying topological defects by
showing for instance that a generalisation of the string properties in
an arbitrary number of dimensions is possible, which in turn give a
new understanding of these string features. With this idea in mind,
we end this article by a derivation of the divergent behaviour
of the timelike component of the current as a function of the
topological defect internal dimension.

\section{Wall model}

Domain walls form whenever a discrete symmetry is spontaneously
broken. The simplest way to achieve that is to break a $Z_2$ symmetry
by means of a scalar $\phi$ whose vacuum expectation value shall be
taken as $\langle 0 |\varphi |0\rangle = \pm \eta$, with $\eta$ the
energy scale of symmetry breaking. This Higgs field may be coupled
with hypercharge-carrying fields which we approximate~\cite{witten} by
a single complex scalar field $\Sigma$ whose vacuum dynamics we require
to be invariant under some $U(1)$ phase transformation group.
In much the same way as
was done for current-carrying cosmic strings~\cite{neutral,enon0}, we
neglect any long range interaction and thus assume a global $U(1)$
symmetry~\cite{neutral}, thereby emphasizing on the actual dynamics of
the wall, assuming charge-coupling corrections to be
negligible, as was shown to be the case for superconducting cosmic
strings~\cite{enon0}. The Lagrangian we shall start with is
therefore
\begin{equation} {\cal L} = -{1\over 2} |\partial_\mu \varphi |^2 -
{1\over 2} |\partial_\mu \Sigma |^2
-V(\varphi,\Sigma),\label{lag}\end{equation}
with the general interaction potential given by
\begin{equation} V(\varphi,\Sigma)={\lambda_\phi \over 8}(\varphi ^2
-\eta^2)^2 + f|\Sigma |^2 (\varphi ^2-\eta^2) +{m_\sigma^2\over 2}
|\Sigma |^2 + {\lambda_\sigma\over 4} |\Sigma |^4 .\label{pot}
\end{equation}
The dynamics given by this Lagrangian include existence of domain
walls, i.e., solutions of the fields equations that separate domains
where $\langle 0 |\varphi |0\rangle = +\eta$ from regions where
$\langle 0 |\varphi |0\rangle = - \eta$, and on which therefore
$\langle 0 |\varphi |0\rangle = 0$. For now on, we shall simply write
$\varphi$ for $\langle 0 |\varphi |0\rangle$. The wall solution will
be a stationnary solution, with the wall locally identified with the
$(x,y)$ plane, the various field amplitudes depending only on the
third $z$ coordinate. Our ansatz is thus
\begin{equation} \varphi = \varphi (z) \ \ \ \hbox{and} \ \ \ \Sigma =
\sigma (z) \exp [i(kx-\omega t)],\label{ansatz}\end{equation}
where we have chosen the frame where the spacelike component of the
current, defined below, is directed along the $x$ direction [this
form~(\ref{ansatz}) for $\Sigma$ can always be attained locally
by means of a simple rotation in the wall plane]. The conserved current,
derived as the Noether invariant under phase transformations, is
\begin{equation} J_\mu = {i\over 2}{\delta {\cal L}\over \delta \partial
^\mu \Sigma } \Sigma^* + \hbox{c.c.} = {i\over 2}\Sigma \stackrel
{\leftrightarrow}{\partial_\mu} \Sigma^* ,\end{equation}
which, with Eq.~(\ref{ansatz}) plugged in yields
\begin{equation} J_\mu = (k\delta_{\mu x}-\omega \delta_{\mu t})
\sigma ^2 (z). \label{current}\end{equation}
The field equations derived from the Lagrangian~(\ref{lag}) under the
assumptions~(\ref{ansatz}) read
\begin{equation} {d^2\varphi \over dz^2}=[{\lambda_\phi\over 2}
(\varphi^2-\eta^2)+2f\sigma^2 ]\varphi,\label{phieq}\end{equation}
\begin{equation} {d^2\sigma \over dz^2}=[w+2f(\varphi^2-\eta^2)+
m_\sigma^2+\lambda_\sigma\sigma^2 ]\sigma,\label{sigeq}\end{equation}
in which we have defined the state parameter
\begin{equation} w\equiv k^2 -\omega^2,\label{state}\end{equation}
whose sign reflects the spacelike or timelike character of the current
given above by Eq.~(\ref{current}) since
\begin{equation} J_\mu J^\mu = w \sigma^4 (z),\end{equation}
and in the chosen conventions of Eq.~(\ref{lag}), the Minkowski metric
is $\eta^{\mu\nu}=$~Diag~$\{ -1,1,1,1 \}$.

The possibility of a current in the wall can be seen in two ways.
First, one can notice that the minimum of the potential, in the actual
vacuum, is given by
\begin{equation} \varphi =\pm\eta \ \ \ \hbox{and} \ \ \ \Sigma =0 ,
\end{equation}
and that this minimum is shifted in the wall where $\varphi=0$ to
\begin{equation} \lambda_\sigma |\Sigma |^2 = 2f\eta^2 - m_\sigma^2,
\end{equation}
so a condensate may exist provided
\begin{equation} m_\sigma ^2 \leq 2f\eta^2 .\label{cond}\end{equation}
Another way to realize that a condensate will in fact
appear~\cite{witten} in the wall consists in first assuming no
condensate ($\Sigma =0$), and solve the perturbative equation for
$\Sigma$ in the domain wall background.  For $\Sigma=0$, the solution
of Eq.~(\ref{phieq}) is known:
\begin{equation} \varphi = \eta \tanh ({1\over 2}\sqrt{\lambda_\phi}\eta
z),\label{th}\end{equation}
and setting a perturbation in the form $\Sigma = \sigma (z) e^{i\omega
t}$ into Eq.~(\ref{sigeq}) yields the one-dimensional Shr\"odinger
equation for $\sigma$
\begin{equation} -{d^2\sigma\over dz^2} + {\cal V}(z) \sigma = \omega^2
\sigma,\label{schr}\end{equation}
where the potential
\begin{equation} {\cal V}(z)\equiv -2f\eta^2 [1-\tanh ^2 ({1\over 2}
\sqrt{\lambda_\phi}\eta z)] + m^2_\sigma ,\label{potbis}\end{equation}
is negative definite when the condition~(\ref{cond}) holds. Hence,
under this condition, $\Sigma$ evolves in an attractive potential
well, with negative eigenvalues for $\omega^2$. Therefore, there
exists unstable modes and a condensate forms.

\section{Current quenching and phase frequency threshold}

In order to analyse the internal structure of such a current-carrying
domain wall, it turns out to be convenient to introduce a set of
dimensionless functions and variables $\zeta$, $X$, $Y$, $\tilde w$
and $\{ \alpha_{1,2,3} \} $ as
\begin{equation} \varphi (z) = \eta X(\zeta ) ,\end{equation}
\begin{equation} \sigma (z) = {m_\sigma \over \sqrt{\lambda_\sigma}}
Y( \zeta ),\end{equation}
with
\begin{equation} \zeta = \sqrt{\lambda_\phi} \eta z.\end{equation}
The state parameter is similarly rescalled into
\begin{equation} w = {\lambda_\phi \lambda_\sigma \eta^4 \over
m_\sigma^2} \tilde w,\label{rescw}\end{equation} and provided we
redefine the arbitrary underlying parameters as~\cite{neutral,enon0}
\begin{equation}\alpha_1 = {m_\sigma^2\over\lambda_\sigma\eta^2} \ \ \ ,
\ \ \ \alpha_2 = {fm_\sigma^2\over\lambda_\phi\lambda_\sigma\eta^2} \ \
\ \hbox{and} \ \ \ \alpha_3 = {m_\sigma^4\over\lambda_\phi
\lambda_\sigma\eta^4},\end{equation}
we get the very simple set of ordinary differential equations
\begin{equation} X'' = X [{1\over 2}(X^2-1)+2\alpha_2 Y^2],
\label{X}\end{equation}
\begin{equation}  \alpha_1 Y'' = Y [\tilde w + 2\alpha_2 (X^2-1)
+\alpha_3
(Y^2+1) ],\label{Y}\end{equation}
where a prime denotes a derivative with respect to $\zeta$.

Two constraints on these parameters arise from the requirement that
the theory be physically meaningful and consistent with currents
flowing along the wall. The condition~(\ref{cond}) for instance, reads
in terms of these parameters
\begin{equation} \alpha_3 \leq 2 \alpha_2 ,\label{cd1}\end{equation}
while demanding that the energy of the wall configuration ($\varphi
=0$ and $\Sigma \not= 0$) be greater than the actual surrounding
vacuum configuration
($\varphi =\eta$ and $\Sigma = 0$) implies
\begin{equation} (\alpha_3-2\alpha_2)^2 \leq {1\over 2}\alpha_3
\label{cd2}.\end{equation}

The first of these constraints in fact means that there exists a
spacelike saturation current which cannot be exceeded. To see
that this is indeed the case, let us perform an expansion of
$X$ and $Y$ close to
the wall where $\zeta\ll 1$, in the form~\cite{neutral}
\begin{equation} X \sim x_1 \zeta + b \zeta^3 \ \ \ \hbox{and} \ \ \
Y\sim y_0 - a \zeta ^2,\label{near0}\end{equation} which satisfy the
boundary conditions on the wall worldsheet, and in particular
regularity of the $\Sigma$ field [which accounts for $Y' (0)=0$].
Plugging back into Eqs.~(\ref{X}) and~(\ref{Y}) gives
\begin{equation} b = x_1 (2\alpha_2 y_0^2 -{1\over 2}),\end{equation}
and
\begin{equation} a = {y_0\over 2\alpha_2} [2\alpha_2 -\tilde w -
\alpha_3 (y_0^2 +1) ],\label{27}\end{equation}
so that because the condensate is actually at its maximum at $z=0$,
one has $a\geq 0$, which means
\begin{equation} \tilde w \leq 2\alpha_2-\alpha_3 .\end{equation}
Thanks to the requirement~(\ref{cd1}), we see that the limit applies
only in the spacelike current case where $\tilde w\geq 0$, and it
reflects the fact that for $\tilde w = 2\alpha_2-\alpha_3$, one has
$y_0=0$, and therefore no condensate, hence no current. So there exist
a value of the state parameter above which the current quenches to
zero.

On the other hand, investigating the large $\zeta$ behaviour of
Eqs.~(\ref{X}) and~(\ref{Y}) yields the following asymptotics
\begin{equation} 1-X\sim \exp (-\zeta ),\end{equation}
as expected from the knowledge of the true solution~(\ref{th}) in the
decoupled case ($X_{\alpha_2 =0}\sim 1- 2e^{\zeta}$), and
\begin{equation} Y\sim \exp (-\sqrt{{\tilde w +\alpha_3\over \alpha_1}}
\zeta ),\end{equation}
for positive $\tilde w +\alpha_3$,
\begin{equation} Y\sim \cos (\sqrt{|{\tilde w +\alpha_3\over \alpha_1}|}
\zeta +\delta ),\end{equation}
for negative $\tilde w +\alpha_3$, with the special $\tilde w
=-\alpha_3$ case leading to
\begin{equation} Y\sim \sqrt{2\alpha_1 \over \alpha_3} \zeta^{-1}.
\end{equation}
Thus, exactly as in the case of a current carrying cosmic string,
there exists a phase frequency threshold given by $\tilde w =
-\alpha_3$, or $\omega = m_\sigma$, above which the integral of the
current~(\ref{current}) from the sheet to infinity diverges. This is
therefore not a mechanism depending on the dimension of the
topological defect under consideration, and can be interpreted as
charge carriers evaporation from it~\cite{neutral}. This phase frequency
threshold is discussed more thoroughly at the end of the following
section where integrated quantities are explicitely calculated.

\section{Macroscopic quantities}

For most of the cosmologically relevant calculations with topological
defects, it is convenient to consider them as infinitely thin, and for
that purpose, it is necessary to know the stress energy tensor and the
current as line integrals starting from the wall's worldsheet to
infinity. For instance, the integrated current reads
\begin{equation} {\cal C}\equiv 2\int dz \sqrt{|J_\mu J^\mu|} = 2
\sqrt{|w|}\int dz \sigma^2 (z) = 2\eta^2\sqrt{\alpha_1} |\tilde \nu|
\int d\zeta Y^2(\zeta),\label{Cint}\end{equation}
where we have defined $\nu=~$Sign~$(w)\sqrt{|w|}$ and rescalled it
according to Eq.~(\ref{rescw}); the additional factor of 2 is here to
account for both sides of the wall. The parameter $\nu$, being essentially
identifiable as $k$ or $-\omega$, is readily interpreted and has thus
been used as the relevant parameter for the plots presented below.

Another obviously very useful quantity for a macroscopic description
of a surface current-carrying domain wall is its stress energy tensor
\begin{equation} T^{\mu\nu}=-2g^{\mu\alpha}g^{\nu\beta}{\delta {\cal L}
\over \delta g^{\alpha\beta}}+g^{\mu\nu}{\cal L},\end{equation}
which, in the case under consideration, needs to be diagonalized.
It is worth noting at this point that even though the existence of
a current in the wall indeed breaks the Lorentz invariance along the
worldsheet, thereby raising the stress-energy tensor's degeneracy, it
does so through the introduction of one privileged direction. Hence, just
like in the string's case, there can be only two different eigenvalues,
namely the energy per unit area $U$, and the surface tension $T$. The
resulting stress-energy tensor then reads
\begin{equation} {\bf T}_{(<0)} \equiv \left(
\matrix{U &    &    &   \cr
          & -T &    &   \cr
          &    & -T &   \cr
          &    &    & 0 \cr}\right),\label{Tneg}\end{equation}
for a timelike current (for which the spatial isotropy is left
unbroken), whereas the spacelike current case similarly yields
\begin{equation} {\bf T}_{(>0)} \equiv \left(
\matrix{U &    &    &   \cr
          & -U &    &   \cr
          &    & -T &   \cr
          &    &    & 0 \cr}\right).\label{Tpos}\end{equation}

We shall now calculate explicitely these eigenvalues in the specific
case~(\ref{lag}) under consideration, and for that purpose, we perform
a Lorentz boost in the $x-$direction in such a way that the phase of the
current carrier $\Sigma$ reads $kz$ or $-\omega t$. In this frame, in
which we shall for now on remain except when it comes to the lightlike
case, one has the energy per unit surface
\begin{equation} U=2\int dz T_{tt}=\sqrt{\lambda_\phi}\eta^3 \int d\zeta
\left[ X'^2 +\alpha_1 Y'^2 +|\tilde w|Y^2 +{1\over 4}(X^2-1)^2 +2\alpha_2
Y^2(X^2-1)+\alpha^3 Y^2({1\over 2}Y^2+1)\right],\label{U}\end{equation}
the surface tension parallel to the current
\begin{equation} T_{\|}=-2\int dz T_{xx}=\sqrt{\lambda_\phi}\eta^3
\int d\zeta \left[ X'^2 +\alpha_1 Y'^2 -|\tilde w|Y^2
+{1\over 4}(X^2-1)^2 +2\alpha_2 Y^2(X^2-1)+\alpha^3 Y^2({1\over 2}
Y^2+1)\right],\label{Tx}\end{equation}
the surface tension orthogonal to the current
\begin{equation} T_{\perp}=-2\int dz T_{yy}=\sqrt{\lambda_\phi}\eta^3
\int d\zeta \left[ X'^2 +\alpha_1 Y'^2 +\tilde w Y^2 +{1\over 4}(X^2-1)^2
+2\alpha_2 Y^2(X^2-1)+\alpha^3 Y^2({1\over
2}Y^2+1)\right],\label{Ty}\end{equation} while the last integrated
component provides a very useful numerical constraint as we shall see
shortly because
\begin{equation} T_z=-2\int dz T_{zz}=\sqrt{\lambda_\phi}\eta^3 \int d\zeta
\left[ -X'^2 -\alpha_1 Y'^2 +\tilde w Y^2 +{1\over 4}(X^2-1)^2 +2\alpha_2
Y^2(X^2-1)+\alpha^3 Y^2({1\over 2}Y^2+1)\right]\label{Tz}\end{equation}
should in fact vanish identically. This can be checked almost
immediately when no condensate is present since in that case, one has
$X_0=\tanh \zeta /2$, so that $X'_0=-{1\over 2}(X^2 -1)$ which in turn
implies
\begin{equation} T^{(0)}_z=\sqrt{\lambda_\phi}\eta^3 \int d\zeta \left[
-X'^2 +{1\over 4}(X^2-1)^2 \right] =0,\end{equation}
while the general case gives, with the ansatz~(\ref{ansatz})
$$\partial_x T^{xx}=\partial_y T^{yy}=\partial_t T^{tt}
=0$$ and finally, conservation of the stress
energy tensor $\partial_\mu T^{\mu\nu}=0$ yields
\begin{equation}\partial_z T^{zz}= 0.\label{numb}\end{equation}
But the boundary conditions one must use are such that asymptotically,
the fields take their vacuum values, so $$\lim_{z\to\infty} \sigma (z) =
\lim_{z\to\infty} \partial_z \sigma (z) = \lim_{z\to\infty} \partial_z
\varphi (z) = \lim_{z\to\infty} (\varphi^2 -\eta^2) = 0,$$
so $\lim_{z\to\infty} T^{zz}=0$ which, with Eq.~(\ref{numb}) implies
$T^{zz}=0$. Hence, Eq.~(\ref{Tz}) provides a constraint on the fields,
namely
\begin{equation} X'^2+\alpha_1 Y'^2 = \tilde w Y^2 +{1\over 4} (X^2-1)^2
+2\alpha_2 Y^2 (X^2-1) +\alpha_3 Y^2 ({1\over 2}Y^2+1),\label{*}
\end{equation}
which is used for numerical purposes since it gives the value of the
derivative of $X$ near the origin, i.e. $x_1$ with the notation of
Eq.~(\ref{near0}), as a function of the $\Sigma$ field's VEV $y_0$,
with
\begin{equation}x_1^2={1\over 4} + y_0^2 [\alpha_3 (1+2y_0^2) -2\alpha_2
-\tilde w].\label{**}\end{equation}
Note first that we recover
$x_1^2=1/4$ in the noncurrent carrying case, again in agreement with
the corresponding known analytic solution, and second that
Eq.~(\ref{*}) is not a trivial constraint: as numerical integration
reveals, the functional $U[X(\zeta),Y(\zeta)]$ has two extrema
depending on the field configuration, one of which corresponds to an
unphysical maximum, whereas the second is indeed a minimum satisfying
Eq.~(\ref{*}). The numerical program developped for solving
Eqs.~(\ref{X}) and~(\ref{Y}) used therefore the constraint~(\ref{*})
by fixing the parameters at the origin with Eq.~(\ref{**}). Two
criteria for ensuring the convergence to the actual physical solution
were thus considered, namely that the solution should be one indeed
and therefore should extremise $U$, and the vanishing of $T_z$.

A last consideration permits an evaluation of the accuracy of the
numerical results thereby obtained, and it is the final point on the
$\nu$ line calculated for a spacelike current. This point corresponds
to $\tilde w=2\alpha_2-\alpha_3$ which, according to Eq.~(\ref{27})
and the discussion following it, has no current at all. In that case,
all the integrals of Eqs.~(\ref{U},\ref{Tx},\ref{Ty}) are equal to
$U_0$, with
\begin{equation} U_0=\sqrt{\lambda_\phi} \eta^3 \int d\zeta \left[
X_0'^2+{1\over 4}(X_0^2-1)\right]=2\sqrt{\lambda_\phi} \eta^3 \int
d\zeta X_0'^2,\end{equation}
when one takes the solution $X_0=\tanh \zeta /2$, and this is
\begin{equation} U_0=2\sqrt{\lambda_\phi} \eta^3 \int X' dX =-
\sqrt{\lambda_\phi} \eta^3 \int_0^1 (X^2-1)dX = {2\over 3}
\sqrt{\lambda_\phi} \eta^3.\end{equation}
The condensate therefore must respect
\begin{equation} {U_\sigma \over \sqrt{\lambda_\phi} \eta^3 } \leq
{2\over 3}\end{equation} in order to be stable against charge carrier
evaporation, with the equality obtained in the limit $\tilde w\to
2\alpha_2 -\alpha_3$. This in fact also limits the range of variation
of $w$ for a timelike current for it seems doubtful that a state having
$U_\sigma > U_0$ could survive in practice.

The case of a lightlike current shares with the noncurrent-carrying
wall the property that the stress energy tensor's eigenvalues are
strictly equal. It can usually be set, after diagonalization for
$J_\mu J^\mu\not= 0$, as
\begin{equation} T^{\mu\nu}=Uu^\mu u^\nu -T_\| x^\mu x^\nu - T_\perp
y^\mu y^\nu ,\end{equation} with $u^\mu$ the timelike eigenvector
($u_\mu u^\mu =-1$) and $x^\mu$, $y^\mu$ the spacelike eigenvectors
($x_\mu x^\mu =y_\mu y^\mu =1$, and $x_\mu y^\mu =0$), but for a
lightlike current, it reads
\begin{equation} T^{\mu\nu}=U u^\mu u^\nu - T_\| x^\mu x^\nu - T_\perp
y^\mu y^\nu -{1\over 2} (u^\mu x^\nu + u^\nu x^\mu )\omega^2 \int dz
\sigma^2 (z),\end{equation}
where we have set $\Sigma =\sigma (z)\exp[i\omega (t-x)]$. Upon
diagonalization, this reads
\begin{equation} T^{\mu\nu}=T_\| (v^\mu _- v^\nu _- - v^\mu_+ v^\nu_+
-y^\mu y^\nu),\end{equation} with $v^\mu_\pm = {1\over 2}(x^\mu\pm
u^\mu)$ the lightlike eigenvectors of $T^{\mu\nu}$, and $T_\|$ as
given by Eq.~(\ref{Ty}) with $\tilde w =0$.

Let us investigate more thoroughly the spacelike and timelike cases.
The timelike case is characterised, as exemplified on Eq.~(\ref{Tneg}),
by the isotropy of
the purely spatial part of $T^{\mu\nu}$, i.e. $T_\| = T_\perp \equiv
T$. As in the string case, one has the Legendre-like relation
\begin{equation} U-T = -\nu {\cal C}, \ \ \ \nu \leq 0
\label{Leg}\end{equation}
and the now standard formalism developped by Carter~\cite{formal}
applies straightforwardly. The case of a spacelike current is slightly
more involved and perhaps requires more thought for each particular
cosmologically interesting configuration studied because the
spatial isotropy
of the surface is no longer present since the current picks a
privileged spatial direction in the worldsheet. However,
Eqs.~(\ref{Tpos}), (\ref{U}) and~(\ref{Ty}) show that yet another
simplification
arises from the fact that $U=T_\perp$, i.e. the purely spatial
component of the stress energy tensor in the direction parallel to the
current flow is the energy per unit surface. Setting $T= T_\|$, a
relation similar to Eq.~(\ref{Leg}) is obtained in the form
\begin{equation} U-T= \nu {\cal C},\ \ \ \nu \geq 0
\label{Leg2}\end{equation}
which can be understood in terms of duality between spacelike and
timelike currents~\cite{formal}. The relevant rescalled integrals are
displayed on the figures.

Fig.~1 represents the energy per unit area
and the surface tensions as functions of the rescalled state parameter
$\tilde\nu$
for a specific set of parameters $\{ \alpha_i\}$ (chosen to yield
a generic kind of result as well as giving measurable effects), with
Fig.~1.a showing the variations of $U(\tilde\nu )$ and $T
(\tilde\nu )$ for a spacelike current-carrying wall having a positive
state parameter $\tilde\nu >0$, while Fig.~1.b represents
$U(\tilde\nu )$ and $T(\tilde\nu )$ for a timelike current-carrying
wall with $\tilde\nu >0$. Similarly, Figs.~2.a and~2.b show the
amplitude of the current~(\ref{current}) in the magnetic and electric
regimes respectively. As might have been anticipated, these figure are
very much like those obtained for a neutral current-carrying cosmic
string~\cite{neutral}, at least in the  classically stable part of
the equation of state, which is definable through the requirement
that the soundlike perturbation squared velocity $c_{L}^2=-dT/dU$
be positive. Thus, the approximate analytic equation of state proposed
in Ref.~\cite{analytic} should be useful also in this domain wall
context. In fact, the only noticable difference between the wall and
the string as far as internal structure goes concerns the unstable
region: Figs.~1.a and~2.a shows that almost as soon as the wall
becomes unstable with respect to soundlike perturbations along the
wordlsheet, the wall's stress energy tensor tends to the ordinary wall
one, namely the isotropic stress tensor with unique eigenvalue
$U_0 = {2\over 3} \sqrt{\lambda_\phi}\eta^3$. Therefore, most of
the current-carrying domain wall properties are essentially similar
to the string properties.

Finally, let us remark the following important mathematical property
of the surface current-carrying domain wall. As is the case for a
superconducting cosmic string, it can be seen that there exists a
phase frequency threshold~\cite{neutral} given by $w = -m^2_\sigma$
at which point the current~(\ref{Cint}) diverges. For the cosmic string
case, the first order pole behaviour ${\cal C}_{String}
\sim (w+m^2_\sigma)^{-1}$
was found~\cite{neutral} whereas the wall case yields
${\cal C}_{Wall}\sim (w+m^2_\sigma)^{-1/2}$. This is because in
both cases, denoting by $d$ the codimension of the topological
defect, i.e., 2 for a string and 1 for a wall in a 4 dimensionnal
background, the current carrier field is seen to satisfy Eq.~(\ref{sigeq})
which, far from the topological defect, gives the relation
\begin{equation} \Delta _d \sigma \sim (w+m^2_\sigma) \sigma ,
\label{asymp}\end{equation}
where $\Delta _d$ stands for the Laplacian in $d$ dimensions: this
is simply $d^2/dz^2$ in the wall case under consideration here, and
$d^2/dr^2 +{d-1\over r}d/dr$ in the general case with $r$ the ``radial''
distance to the defect's core. Setting $\chi = kr$, with $r\equiv
z$ in our wall case and $k^2  = w+m^2_\sigma$, one can extract
$\sigma$ as a function of $k$ since for $k\not= 0$,
Eq.~(\ref{sigeq}) [i.e., Eq.~(\ref{asymp})] transforms into
\begin{equation} {d^2\sigma \over d\chi ^2} + {d-1\over \chi}
{d\sigma\over d\chi} = \sigma (\chi ), \label{genBessel}\end{equation}
whose solution cannot depend on $k$

The solution to Eq.~(\ref{genBessel}) is well known:
\begin{equation} \sigma \sim A \chi^{1-d/2} K_0 (\chi )
,\label{solas}\end{equation}
with $K_0$ the Bessel function of zeroth order whose asymptotic
behaviour is given by $K_0 (\chi ) \sim \exp (-\chi) /\sqrt{\chi}$.
Thus, one finds the general phase frequency threshold behaviour, up
to a finite part [corresponding to the fact that one has to integrate
up to the point where the approximation~(\ref{asymp}) becomes valid]
\begin{eqnarray} {\cal C} &\propto &\int r^{d-1} dr \sigma ^2 (k
r)\nonumber \\ &\propto & {1\over k^d}\int \chi K_0^2 (\chi )
d\chi \nonumber\\ &\propto & (w+m^2_\sigma)^{-d/2} ,
\end{eqnarray}
with $d=1$ for a current-carrying domain wall, $d=2$ for a
superconducting cosmic string, and $d=3$ for a charged monopole
in a four dimensional background spacetime. It is in fact possible
to be slightly more precise concerning this divergence: the function
$\kappa$, defined as~\cite{formal,witten}
\begin{equation} \kappa\equiv 2 {dU\over dw} = 2 \int d^d {\bf x}_\perp
\sigma^2 ({\bf x}_\perp), \end{equation}
being proportionnal to ${\cal C}$, also diverges, and it may be
seen that, under the assumption that $y_0^2\sim {2\alpha_2\over
\alpha_3}$ near the threshold [see Ref.~\cite{neutral} and
Eq.~(\ref{27})]
\begin{equation} \kappa = \kappa_f (w) + A {f\eta^2\over\lambda_\sigma}
(w+m^2_\sigma)^{-d/2},\end{equation}
which is valid for various values of the codimension $d$, with $\kappa_f
(w)$ the finite part of $\kappa$ and $A$ a pure number, calculable
in principle by a matching of the asymptotic solution~(\ref{solas}) to
the origin and depending on $d$. Note that the dimension of this
function $\kappa$ is given straightforwardly once $d$ is known.

\section*{Acknowledgments}

I wish to thank L.~Blanchet, B.~Carter, E.~Copeland,
A.-C.~Davis and T.~Vachaspati for many stimulating
discussions. I would also like to thank the Isaac
Newton Institute for Mathematical Sciences in
Cambridge (UK) for their hospitality during the
time this work was being done.

\figure{Energy per unit area $U$ (solid line) and surface tensions
$T_{>0}=T_{\|}$ and $T_{<0}=T_{\perp}$ (dashed lines) as functions
of the rescalled state parameter $\tilde \nu$ and in units of
$\sqrt{\lambda_\phi}\eta^3$. Fig.~1.a
represents the equation of state in the timelike case $\tilde \nu
\leq 0$, while Fig.~1.b is for the spacelike range $\tilde \nu \geq 0$.}

\figure{Integrated value of the surface current in units of $\eta^2$ for
the same variation ranges as on Fig.~1 as a function of $\tilde\nu$.}

\end{document}